# Many-body localization with mobility edges

Yichen Huang (黄溢辰)*
*Department of Physics, University of California, Berkeley, Berkeley, California 94720, USA*
(Dated: July 6, 2015)

We construct a solvable spin chain model of many-body localization (MBL) with a tunable mobility edge. This simple model not only demonstrates analytically the existence of mobility edges in interacting one-dimensional (1D) disordered systems, but also allows us to study their physics. By establishing a connection between MBL and a quantum central limit theorem (QCLT), we show that many-body localization-delocalization transitions can be visualized as tuning a mobility edge in the energy spectrum. Since the effective disorder strength for individual eigenstates depends on energy density, we identify "energy-resolved disorder strength" as a physical mechanism for the appearance of mobility edges, and support the universality of this mechanism by arguing its presence in a large class of models including the random-field Heisenberg chain. We also construct models with multiple mobility edges. All our constructions can be made translationally invariant.

In one- and two-dimensional disordered free(-fermion) systems, Anderson localization says that all single-particle states are localized, and thus rules out the existence of a mobility edge in the energy spectrum separating localized and delocalized states [2]. As an active area of research, MBL studies the effects of interactions added to an Anderson insulator [4, 10, 14, 19–21, 25]. An important problem is whether mobility edges can exist in 1D disordered systems in the presence of interactions. If they do exist, what is the physics of mobility edges?

Recently, numerical progress has been made in favor of the existence of mobility edges [16–18], although there are still different opinions in the community [7]. To date, exact diagonalization (ED) is the only numerical method for static properties of quantum many-body systems at finite energy density, but it is limited to small system sizes for the simple reason that the dimension of the Hilbert space grows exponentially with the system size. In particular, the state-of-the-art ED is able to work with Hilbert spaces of dimension 705432 or $\lesssim 20$ spin-1/2's [17], and finite-size effects are not always negligible. (The time-evolving block decimation algorithm [31] can efficiently simulate the dynamics of 1D MBL systems [3, 29, 35] due to the slow growth of entanglement [1, 13, 27, 32, 33].)

We construct a "solvable" spin chain model of MBL with a tunable mobility edge. Our approach is fully analytical and (almost) rigorous so that there should be no confusion about the existence of mobility edges in interacting 1D disordered systems. Furthermore, this simple model allows us to study the physics of mobility edges. By establishing a connection between MBL and QCLT, we show that many-body localization-delocalization transitions can be visualized as tuning a mobility edge in the energy spectrum. Although the disorder strength in the Hamiltonian is fixed, the effective disorder strength (to be defined precisely) for individual eigenstates depends on energy density. This motivates us to identify "energy-resolved disorder strength" as a physical mechanism for the appearance of mobility edges, and we support the universality of this mechanism by arguing its presence in a large class of models including the random-field Heisenberg chain. We also construct models with two mobility edges such that one region of localized (delocalized) states sandwiches two regions of delocalized (localized) states. All our constructions can be made translationally invariant.

*Physical picture.*—Before presenting mathematical details, we discuss intuitions and essential ideas at a non-rigorous level. To be specific, we consider the random-field Heisenberg chain

$$H(\{h_i\}_{i\in \mathbf{Z}}) = \sum_{i\in \mathbf{Z}} \vec{S}_i \cdot \vec{S}_{i+1} - h_i S_i^z, \qquad (1)$$

where $\vec{S}_i = (S_i^x, S_i^y, S_i^z)$ is the spin-$S$ operator at the site $i$, and $h_i$'s are independent and identically distributed (i.i.d.) uniform random variables on the interval $[-h, h]$. We have two different arguments for the existence of mobility edges in this model.

*Argument 1.*—Since the ground-state energy and bandwidth of (1) depend on $\{h_i\}$, we take the union of all eigenvalues and eigenstates for all disorder realizations:

$$\bigcup_{\{h_i\}\in [-h,h]^{\otimes \mathbf{Z}}} \{(E, |\psi\rangle) : H(\{h_i\})|\psi\rangle = E|\psi\rangle\}, \qquad (2)$$

where $H(\{h_i\})$ is called the parent Hamiltonian of $|\psi\rangle$. We argue that the states in (2) with larger (in absolute value) energy densities are more likely to be localized.

Intuitively, different disorder realizations have different disorder strength, and we roughly quantify the disorder strength in each individual disorder realization $\{h_i\}$ by $\sum_i |h_i|/n$, where $n$ is the system size. Admittedly, this is not a faithful measure of disorder strength for some disorder realizations (e.g., $h_i = h$ for $\forall i \in \mathbf{Z}$), but such cases are rare. We observe that the parent Hamiltonians for the states in (2) with large (in absolute value) energy densities must have large disorder strength. For example, we fix $S = 1/2$ and $h = 2.5$. (It is shown numerically that (1) with $S = 1/2$ has mobility edges for $1.5 \lesssim h \lesssim 3.5$; see Figure 1 in [17].) Then, a state with $E/n \leq -1$ implies $\sum_i |h_i|/n \geq 2 + 2E_0 \approx 1.11$, where $E_0 = 1/4 - \ln 2$ is the ground-state energy density of the (homogeneous) spin-1/2 anti-ferromagnetic Heisenberg chain.

* yichenhuang@berkeley.edu

*Argument 2.*—We argue that for a single disorder realization of (1) with $S \geq 1$, the eigenstates at larger (in absolute value) energy densities are more likely to be localized. We use the excited-state real-space renormalization group (RSRG-X) technique, which is a heuristic approach to solve or make progress towards solving the eigenvalues and eigenstates in strongly disordered systems [12, 23, 30]. As any other renormalization group procedure, RSRG-X sequentially "integrates out" some degrees of freedom, while generating effective interactions between the remaining degrees of freedom. Thus, the disorder strength is modified. As the implementation of RSRG-X depends on the energy of the targeting eigenstate, the effective Hamiltonians for eigenstates at different energy densities may have different disorder strength.

We do one layer of RSRG-X for (1). Assuming $h \gg 1$, we identify a set $T \subset \mathbf{Z}$ such that (i) $|h_i| \gg \max\{|h_{i-1}|, |h_{i+1}|, 1\}$ for $\forall i \in T$; (ii) $|T|$ is of the order of the system size. Fix $S = 1$. For each $i \in T$, (i) implies that the spin $i$ is (approximately) in an eigenstate $|s_i^z = 0, \pm 1\rangle$ of $S_i^z$. Then a perturbative calculation shows that in the effective Hamiltonian acting on the remaining spins $\mathbf{Z} \setminus T$, the terms in a neighborhood of $i$ read

$$- (h_{i-1} - s_i^z)S_{i-1}^z - (h_{i+1} - s_i^z)S_{i+1}^z + H_{i-1,i+1}, \quad (3)$$

where $\|H_{i-1,i+1}\| = O(1/|h_i|)$ acts on the spins $i-1$ and $i+1$ so that the spin chain is not completely decoupled. Unlike $s_i^z = 0$, $s_i^z = \pm 1$ induces fluctuations in the random fields and thus increases the disorder strength in the effective Hamiltonian. We expect that $\sum_{i \in T} |s_i^z|$ tends to be larger when the targeting eigenstate of (1) has larger (in absolute value) energy density; see the paragraph containing (13) for a quantitative analysis of a very similar statement.

Since the disorder strength of the parent or effective Hamiltonians for individual eigenstates depends on energy density, we propose "energy-resolved disorder strength" as a physical mechanism for the appearance of mobility edges. Obviously, our arguments apply to not only (1) but also a large class of models, which suggests the universality of this mechanism. We will construct a simple spin chain model (5) of MBL such that the energy-resolved disorder strength can be solved analytically: It decreases continuously and monotonically as energy density increases. This implies a mobility edge below (above) which almost all eigenstates are localized (delocalized).

*Preliminaries.*—To proceed, we need some formal definitions. An (interacting) model is MBL if almost all its eigenstates are localized. In the literature, there are several (inequivalent) criteria for whether an individual state is localized. Our construction does not rely on which criterion to use. For example, you might keep in mind that a state is localized if it satisfies an area law for entanglement [5, 9, 26].

We take a detour and discuss QCLT on the distribution of the eigenvalues of a local Hamiltonian. In our context, it says that all but an exponentially small (in the system size) fraction of eigenstates have the same energy density up to corrections vanishing in the thermodynamic limit. QCLT can be proved for any local Hamiltonian [6, 8, 15]. Here we just prove (a weak version of) it in all weakly interacting systems. Indeed, the single-particle spectrum of any (homogeneous or disordered) local free-fermion Hamiltonian is bounded. For a random many-body eigenstate, its energy is a random sum of single-particle energies, and the (classical) CLT implies that its energy density is close to the mean with overwhelming probability, cf. 1D random walk with random but bounded step size. (Weak) interactions can broaden any energy density interval by at most $\epsilon$ if the norm of the interaction terms per site is upper bounded by $\epsilon$.

The definition of MBL reads "... if almost all ..." because one cannot rule out the possibility that a very small fraction of eigenstates are delocalized (especially in random systems). However, this seemingly innocent definition has a caveat. Even if all but an exponentially small fraction of eigenstates are localized, there can still be an exponential number of delocalized eigenstates. Furthermore, it is possible that almost all eigenstates away from the mean energy density are delocalized or that mobility edges exist. Indeed, QCLT implies that a model is MBL if almost all its eigenstates at the mean energy density are localized.

As we will encounter probability distributions of various shapes, a formal definition of disorder strength appears necessary. Let $X$ be a real-valued random variable with probability density function $p(x)$, where $p(x) \geq 0$ and $\int_{-\infty}^{+\infty} p(x) dx = 1$. We only consider symmetric probability distributions with compact support such that $p(x) = p(-x)$ and $\int_{-\Lambda}^{\Lambda} p(x) dx = 1$ for $\Lambda = O(1)$. The disorder strength $s(X)$ of a random variable is a measure of how broad $p(x)$ distributes. Formally, $s$ is a nonnegative function of random variables satisfying

(i) $s(X) = 0$ if and only if $p(x)$ is a delta function.
(ii) Let $X_k (k = 1, 2)$ be a uniform random variable on the interval $[-\Lambda_k, \Lambda_k]$ with $\Lambda_1 < \Lambda_2$, and $(1-\lambda)X_1 \text{OR} \lambda X_2$ denote a random variable which is $X_2$ with probability $0 \leq \lambda \leq 1$ and $X_1$ otherwise. Then $s((1-\lambda)X_1 \text{OR} \lambda X_2)$ is a continuous monotonically increasing function of $\lambda$.
(iii) Other postulates irrelevant to us.

A canonical example of $s$ is the variance of a random variable, but this may not be a faithful measure of disorder strength from a physical point of view, e.g., in the random-field Heisenberg chain (1), two different distributions of $h_i$'s with the same variance may correspond to the different phases. Hence, we do not specify $s$, but only assume that a faithful measure of disorder strength exists for the random spin model (4).

*The model.*—We now present the details of our model.

Similar but not identical constructions appeared previously in different contexts, e.g., [11, 22, 24, 28, 34].

We start with a disordered Hamiltonian on a spin chain (in the thermodynamic limit), where the local dimension $d = \Theta(1)$ of each spin is a small constant,

$$H = \sum_{i \in \mathbf{Z}} H_{i,i+1} + \lambda_i H_i. \qquad (4)$$

Here, $\|H_{i,i+1}\| \leq 1$ is a translationally invariant nearest-neighbor interaction between the spins $i$ and $i+1$; $\|H_i\| \leq 1$ is a translationally invariant on-site term acting on the spin $i$; $\lambda_i$'s are i.i.d. random variables. Without loss of generality, we assume $\operatorname{tr} H_{i,i+1} = \operatorname{tr} H_i = 0$ so that the mean energy density of $H$ is 0.

Suppose $H$ has a many-body localization-delocalization transition tuned by disorder strength: There is a critical $s_* = \Theta(1)$ such that almost all eigenstates of $H$ are localized (delocalized) if $s(\lambda_i) > (<) s_*$. Let $X_k (k = 1, 2)$ be a uniform random variable on the interval $[-\Lambda_k, \Lambda_k]$ with $s(X_1) < s_* < s(X_2)$. As $s((1 - \lambda)X_1 \text{OR} \lambda X_2)$ is a continuous monotonically increasing function of $\lambda$, there exists $0 < \lambda_* < 1$ such that $s_* = s((1 - \lambda_*)X_1 \text{OR} \lambda_* X_2)$. Although $s_*$ is fixed, $\lambda_*$ is tunable in the sense that $\Lambda_2$ is tunable.

Based on (4), we construct a simple spin chain model (5) of MBL with a tunable mobility edge. Here, a mobility edge is an energy density below (above) which almost all eigenstates are localized (delocalized). We have two spins per unit cell (or two particle species in the language of [11, 22, 24, 34]). The main idea of the construction is to have the second spin species control (via the second term in (5)) the disorder strength in the effective Hamiltonian (7), which acts on the first spin species and depends on the state of the second species. Then, we move the eigenstates of (5) whose effective Hamiltonians (7) have strong (weak) disorder to the bottom (top) of the energy spectrum of (5) by adding a uniform field (the third term in (5)) on the second spin species, which results in a negative gradient in the energy-resolved disorder strength.

In each unit cell, the first spin has local dimension $d$, and the second has local dimension 2. You may combine these two spins into a single spin of larger local dimension $2d = \Theta(1)$ if you prefer one spin per unit cell. We label all spins by two indices $(i, j)$. Here, $i \in \mathbf{Z}$ is the unit cell index; $j = 1, 2$ is the species index. The Hamiltonian is

$$\mathbf{H} = \sum_{i \in \mathbf{Z}} \mathbf{H}_{i,1,i+1,1} + \mathbf{H}_{i,1,i,2} + \hat{\sigma}_i^z, \qquad (5)$$

where $\mathbf{H}_{i,j,i',j'}$ acts on the spins $(i, j)$ and $(i', j')$, and $\hat{\sigma}_i^z$ is the spin-1/2 Pauli operator for $(i, 2)$. Specifically, $\mathbf{H}_{i,1,i+1,1} = \epsilon H_{i,i+1}/(1 + \Lambda_2)$ is translationally invariant, and $\mathbf{H}_{i,1,i,2} = \epsilon H_i \otimes \operatorname{diag}\{X_1, X_2\}/(1 + \Lambda_2)$ involves randomness, where $\epsilon$ is a very small constant. (5) is a weakly interacting system as

$$\|\mathbf{H}_{i,1,i+1,1}\| + \|\mathbf{H}_{i,1,i,2}\| \leq \frac{\epsilon}{1 + \Lambda_2} + \frac{\epsilon \Lambda_2}{1 + \Lambda_2} \leq \epsilon. \qquad (6)$$

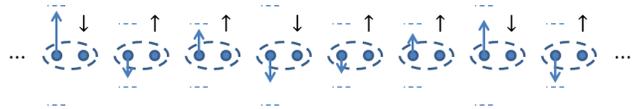

FIG. 1. (Color online) We have two spins (dots) per unit cell (dashed circle). Conditioned that the right spins (second species) are in the state $|\cdots \downarrow\uparrow\uparrow\downarrow\uparrow\uparrow\downarrow\uparrow \cdots\rangle$ (black arrows), the lengths and directions of the blue arrows, respectively, illustrate the magnitudes and signs of the random fields acting on the left spins (first species) in a typical disorder realization of the effective Hamiltonian (7). The short dashed lines show the maximum random-field strength in every unit cell, which is "controlled" by the state of the second spin species.

As $\{\hat{\sigma}_i^z\}$ is a set of conserved quantities, in any eigenstate of $\mathbf{H}$ the spins $(i \in \mathbf{Z}, 2)$ are in a product state. For a particular configuration $\{\sigma_i^z\} \in \{1, -1\}^{\otimes \mathbf{Z}}$, the effective Hamiltonian acting on the first spin species reads

$$\mathbf{H}(\{\sigma_i^z\}) = \sum_{i \in \mathbf{Z}} \frac{\epsilon H_{i,i+1} + \epsilon X_{(3-\sigma_i^z)/2} H_i}{1 + \Lambda_2} + \sigma_i^z, \qquad (7)$$

which is equivalent to (4) after rescaling. Clearly, the second spin species controls the disorder strength in the sense that $X_{(3-\sigma_i^z)/2} = X_1$ if $\sigma_i^z = 1$ and $X_{(3-\sigma_i^z)/2} = X_2$ otherwise. As $\sigma_i^z = \pm 1$ with equal probability in a random eigenstate of $\mathbf{H}$, the (overall) disorder strength in (7) is $s(\frac{1}{2}X_1 \text{OR} \frac{1}{2}X_2)$. Compared with the critical disorder strength $s_* = s((1-\lambda_*)X_1 \text{OR} \lambda_* X_2)$, $\mathbf{H}$ is MBL if we assume $\lambda_* < 1/2$.

Let $z := \sum_i \sigma_i^z / n$ be the average magnetization of the second spin species, where $n$ is the number of unit cells. The energy density (per unit cell) of an eigenstate of $\mathbf{H}$ is well approximated by $z$. Indeed,

$$|\mathbf{H}(\{\sigma_i^z\})/n - z| \leq \frac{\epsilon}{n}\sum_i \frac{\|H_{i,i+1}\| + \Lambda_2\|H_i\|}{1 + \Lambda_2} \leq \epsilon. \qquad (8)$$

Furthermore, QCLT says that for any particular configuration $\{\sigma_i^z\}$ almost all eigenstates of $\mathbf{H}(\{\sigma_i^z\})$ have energy density $z$. As $\sigma_i^z = \pm 1$ with equal probability in a random eigenstate of $\mathbf{H}$, $z$ follows a rescaled Bernoulli distribution, and $|z| = o(1)$ for almost all eigenstates of $\mathbf{H}$. This allows the presence of a mobility edge at energy density $e > 0$.

We now solve the energy-resolved disorder strength in our model by (approximately) identifying $z$ with energy density. When $z$ is fixed,

$$X_{(3-\sigma_i^z)/2} = \frac{1+z}{2}X_1 \text{OR} \frac{1-z}{2}X_2. \qquad (9)$$

Hence $s(X_{(3-\sigma_i^z)/2})$ is a continuous monotonically decreasing function of $z$. Note that by fixing $z$ the random variables $X_{(3-\sigma_i^z)/2}$'s in different unit cells become slightly correlated, but we expect no significant physical effects of this very weak correlation. Compared with the critical disorder strength $s_* = s((1-\lambda_*)X_1 \text{OR} \lambda_* X_2)$, a mobility edge appears at the energy density $e$ such that

$$s(X_{(3-\sigma_i^z)/2})|_{z=e} = s_* \Rightarrow e = 1 - 2\lambda_* > 0, \quad (10)$$

and almost all eigenstates of $\mathbf{H}$ below (above) the energy density $e$ are localized (delocalized).

*Discussions & extensions.*—We have established a simple reduction from a random spin model (4) to another (5) with a nonzero gradient in the energy-resolved disorder strength. If the former has a many-body localization-delocalization transition tuned by disorder strength, then the latter has a mobility edge. Furthermore, we expect that the transition tuned by disorder strength in the former and that tuned by energy density in the latter have the same critical exponents. It should be clear that such a reduction can be established in general. Although we assumed that (4) is a (random) spin chain with only on-site disorder, a minor modification of our construction works in any spatial dimension and applies to cases where the original model has bond disorder.

The model (5) has a mobility edge below (above) which almost all eigenstates are localized (delocalized). Adding a minus sign (trivially) reverses the energy spectrum so that almost all eigenstates above (below) the mobility edge are localized (delocalized). To construct a spin chain model of MBL with two mobility edges, it suffices to modify only the third term in (5):

$$\mathbf{H}_2 = \sum_{i \in \mathbf{Z}} \mathbf{H}_{i,1,i+1,1} + \mathbf{H}_{i,1,i,2} + h_i(I + \hat{\sigma}_i^z)/2, \quad (11)$$

where $h_i = \pm 1$ with equal probability. Note that $\{\hat{\sigma}_i^z\}$ remains a set of conserved quantities, and the effective Hamiltonian acting on the first spin species becomes

$$\mathbf{H}_2(\{\sigma_i^z\}) = \sum_{i \in \mathbf{Z}} \frac{\epsilon H_{i,i+1} + \epsilon X_{(3-\sigma_i^z)/2} H_i}{1 + \Lambda_2} + \frac{h_i(1 + \sigma_i^z)}{2}. \quad (12)$$

Similarly, $\mathbf{H}_2$ is MBL if $\lambda_* < 1/2$.

Indeed, the effect of the third term in (11) is to move the eigenstates of (11) whose effective Hamiltonians (12) have weak disorder to the ends of the energy spectrum of (11), which results in a "bump" in the energy-resolved disorder strength. As before, we (approximately) identify $y := \sum_i h_i(1 + \sigma_i^z)/(2n)$ with energy density. When $y$ is fixed, the probability distribution function of $z$ is

$$p(z) = \frac{C}{2^{zn/2}\left(\frac{(1-z)n}{2}\right)!\left(\frac{(1+z+2y)n}{4}\right)!\left(\frac{(1+z-2y)n}{4}\right)!}, \quad (13)$$

where $C$ is a normalization factor. As $p(z)$ is sharply peaked at $z = y^2$, (9) becomes

$$X_{(3-\sigma_i^z)/2} = \frac{1+y^2}{2}X_1 \text{ OR } \frac{1-y^2}{2}X_2. \quad (14)$$

with overwhelming probability. Compared with the critical disorder strength $s_* = s((1-\lambda_*)X_1 \text{OR} \lambda_* X_2)$, al-most all eigenstates of $\mathbf{H}_2$ at energy densities (in absolute value) $< (>) \sqrt{1 - 2\lambda_*}$ are localized (delocalized). It is also straightforward to construct a random spin chain model with two mobility edges such that one region of delocalized eigenstates sandwiches two regions of localized eigenstates.

All our constructions can be made translationally invariant. We show how to do this using (5) as an example. Recall that $X_k (k = 1, 2)$ is a continuous uniform random variable on the interval $[-\Lambda_k, \Lambda_k]$. Let $X'_k$ be a discrete uniform random variable on the set $S_k = \{x_{k,1}, x_{k,2}, \ldots, x_{k,d'}\}$ such that (i) $d' = O(1)$ is a constant; (ii) $|x_{k,j}| \leq \Lambda_k$ for $j = 1, 2, \ldots, d'$; (iii) $S_k = -S_k$ so that the distribution of $X'_k$ is symmetric; (iv) The numbers in $S_k$ are generic in the sense that (16) has no accidental degeneracy; (v) $s(X_k) = s(X'_k)$ so that

$$\mathbf{H}' = \sum_{i \in \mathbf{Z}} \mathbf{H}_{i,1,i+1,1} + \mathbf{H}'_{i,1,i,2} + \hat{\sigma}_i^z \quad (15)$$

reproduces the physics of (5), where $\mathbf{H}'_{i,1,i,2} = \epsilon H_i \otimes \text{diag}\{X'_1, X'_2\}/(1 + \Lambda_2)$.

We encode all disorder realizations of (15) into a single translationally invariant Hamiltonian (16) by adding a third spin to every unit cell. These spins are of local dimension $d'$, and are labeled by $(i, 3)$ for $i \in \mathbf{Z}$. It suffices to modify only the second term in (15):

$$\mathbf{H}_T = \sum_{i \in \mathbf{Z}} \mathbf{H}_{i,1,i+1,1} + \mathbf{H}'_i + \hat{\sigma}_i^z, \quad (16)$$

where $\mathbf{H}'_i$ is a translationally invariant term acting on all three spins in the unit cell $i$. In the computational basis,

$$\mathbf{H}'_i = \epsilon H_i \otimes \text{diag}\{x_{1,1}, x_{1,2}, \ldots, x_{1,d'}, x_{2,1}, x_{2,2}, \ldots, x_{2,d'}\}. \quad (17)$$

Let $\hat{\tau}_i = \text{diag}\{1, 2, \ldots, d'\}$ be an operator acting on the spin $(i, 3)$. As $\{\hat{\tau}_i\}$ is a set of conserved quantities, in any eigenstate of $\mathbf{H}_T$ the spins $(i \in \mathbf{Z}, 3)$ are in a product state. For a particular configuration $\{\tau_i\} \in \{1, 2, \ldots, d'\}^{\otimes \mathbf{Z}}$, the effective Hamiltonian acting on the first and second spin species is a disorder realization of (15), where the random variable $X'_k$ in the unit cell $i$ takes the value $x_{k, \tau_i}$. Thus, $\mathbf{H}_T$ is a translationally invariant Hamiltonian with a mobility edge.

*Outlook.*—An important open problem is whether the models we constructed are stable against perturbations. We now lack the tools to make progress on this problem. A proof of stability (even merely for a particular model) is notoriously difficult. Numerical methods (especially ED) are usually limited to small system sizes, and may suffer from strong finite-size effects [22].

### ACKNOWLEDGMENTS

The author would like to thank Joel E. Moore for helpful discussions and especially Xie Chen for inspiring ideas that improve the quality and presentation of this paper. This work was supported by NSF DMR-1206515.

[1] E. Altman and R. Vosk. Universal dynamics and renormalization in many-body-localized systems. *Annual Review of Condensed Matter Physics*, 6(1):383–409, 2015.

[2] P. W. Anderson. Absence of diffusion in certain random lattices. *Physical Review*, 109(5):1492–1505, 1958.

[3] J. H. Bardarson, F. Pollmann, and J. E. Moore. Unbounded growth of entanglement in models of many-body localization. *Physical Review Letters*, 109(1):017202, 2012.

[4] D. M. Basko, I. L. Aleiner, and B. L. Altshuler. Metal-insulator transition in a weakly interacting many-electron system with localized single-particle states. *Annals of Physics*, 321(5):1126–1205, 2006.

[5] B. Bauer and C. Nayak. Area laws in a many-body localized state and its implications for topological order. *Journal of Statistical Mechanics: Theory and Experiment*, 2013(09):P09005, 2013.

[6] F. G. S. L. Brandao and M. Cramer. Equivalence of statistical mechanical ensembles for non-critical quantum systems. arXiv:1502.03263.

[7] W. de Roeck, F. Huveneers, M. Muller, and M. Schiulaz. Absence of many-body mobility edges. arXiv:1506.01505.

[8] L. Erdos and D. Schroder. Phase transition in the density of states of quantum spin glasses. *Mathematical Physics, Analysis and Geometry*, 17(3-4):9164, 2014.

[9] M. Friesdorf, A. H. Werner, W. Brown, V. B. Scholz, and J. Eisert. Many-body localization implies that eigenvectors are matrix-product states. *Physical Review Letters*, 114(17):170505, 2015.

[10] I. V. Gornyi, A. D. Mirlin, and D. G. Polyakov. Interacting electrons in disordered wires: Anderson localization and low-$T$ transport. *Physical Review Letters*, 95(20):206603, 2005.

[11] T. Grover and M. P. A. Fisher. Quantum disentangled liquids. *Journal of Statistical Mechanics: Theory and Experiment*, 2014(10):P10010, 2014.

[12] Y. Huang and J. E. Moore. Excited-state entanglement and thermal mutual information in random spin chains. *Physical Review B*, 90(22):220202(R), 2014.

[13] D. A. Huse, R. Nandkishore, and V. Oganesyan. Phenomenology of fully many-body-localized systems. *Physical Review B*, 90(17):174202, 2014.

[14] J. Z. Imbrie. On many-body localization for quantum spin chains. arXiv:1403.7837.

[15] J. P. Keating, N. Linden, and H. J. Wells. Spectra and eigenstates of spin chain Hamiltonians. *Communications in Mathematical Physics*, 338(1):81–102, 2015.

[16] J. A. Kjall, J. H. Bardarson, and F. Pollmann. Many-body localization in a disordered quantum Ising chain. *Physical Review Letters*, 113(10):107204, 2014.

[17] D. J. Luitz, N. Laflorencie, and F. Alet. Many-body localization edge in the random-field Heisenberg chain. *Physical Review B*, 91(8):081103(R), 2015.

[18] I. Mondragon-Shem, A. Pal, T. L. Hughes, and C. R. Laumann. Many-body mobility edge due to symmetry-constrained dynamics and strong interactions. arXiv:1501.03824.

[19] R. Nandkishore and D. A. Huse. Many-body localization and thermalization in quantum statistical mechanics. *Annual Review of Condensed Matter Physics*, 6(1):15–38, 2015.

[20] V. Oganesyan and D. A. Huse. Localization of interacting fermions at high temperature. *Physical Review B*, 75(15):155111, 2007.

[21] A. Pal and D. A. Huse. Many-body localization phase transition. *Physical Review B*, 82(17):174411, 2010.

[22] Z. Papic, E. M. Stoudenmire, and D. A. Abanin. Is many-body localization possible in the absence of disorder? arXiv:1501.00477.

[23] D. Pekker, G. Refael, E. Altman, E. Demler, and V. Oganesyan. Hilbert-glass transition: New universality of temperature-tuned many-body dynamical quantum criticality. *Physical Review X*, 4(1):011052, 2014.

[24] M. Schiulaz and M. Muller. Ideal quantum glass transitions: Many-body localization without quenched disorder. *AIP Conference Proceedings*, 1610:11–23, 2014.

[25] M. Schreiber, S. S. Hodgman, P. Bordia, H. P. Luschen, M. H. Fischer, R. Vosk, E. Altman, U. Schneider, and I. Bloch. Observation of many-body localization of interacting fermions in a quasi-random optical lattice. arXiv:1501.05661.

[26] M. Serbyn, Z. Papic, and D. A. Abanin. Local conservation laws and the structure of the many-body localized states. *Physical Review Letters*, 111(12):127201, 2013.

[27] M. Serbyn, Z. Papic, and D. A. Abanin. Universal slow growth of entanglement in interacting strongly disordered systems. *Physical Review Letters*, 110(26):260601, 2013.

[28] K. Slagle, Z. Bi, Y.-Z. You, and C. Xu. Many-body localization of symmetry protected topological states. arXiv:1505.05147.

[29] R. Vasseur, S. A. Parameswaran, and J. E. Moore. Quantum revivals and many-body localization. *Physical Review B*, 91(14):140202(R), 2015.

[30] R. Vasseur, A. C. Potter, and S. A. Parameswaran. Quantum criticality of hot random spin chains. *Physical Review Letters*, 114(21):217201, 2015.

[31] G. Vidal. Efficient simulation of one-dimensional quantum many-body systems. *Physical Review Letters*, 93(4):040502, 2004.

[32] R. Vosk and E. Altman. Many-body localization in one dimension as a dynamical renormalization group fixed point. *Physical Review Letters*, 110(6):067204, 2013.

[33] R. Vosk and E. Altman. Dynamical quantum phase transitions in random spin chains. *Physical Review Letters*, 112(21):217204, 2014.

[34] N. Y. Yao, C. R. Laumann, J. I. Cirac, M. D. Lukin, and J. E. Moore. Quasi many-body localization in translation invariant systems. arXiv:1410.7407.

[35] M. Znidaric, T. Prosen, and P. Prelovsek. Many-body localization in the Heisenberg $XXZ$ magnet in a random field. *Physical Review B*, 77(6):064426, 2008.